\documentclass[useAMS,usenatbib]{mn2e}

\title{V723 Cas (Nova Cassiopeiae 1995): MERLIN observations from 1996 to 2001}
\author[Heywood et al]
{I. Heywood$^{1,2}$, T. J. O'Brien$^1$, S. P. S. Eyres$^3$, M. F. Bode$^4$, R.J. Davis$^1$\\
$^1$University of Manchester, Jodrell Bank Observatory, Macclesfield, Cheshire SK11 9DL, UK\\
$^2$University of Oxford Astrophysics, Keble Road, Oxford, OX1 3RH, UK\\
$^3$Centre for Astrophysics, University of Central Lancashire, Preston, PR1 2HE, UK\\
$^4$Astrophysics Research Institute, Liverpool John Moores University, Twelve Quays House, Egerton Wharf, Birkenhead, CH41 1LD, UK}

\begin{document}

\date{Accepted 200x month xx. Received 200x month xx}
\pagerange{\pageref{firstpage}--\pageref{lastpage}}
\pubyear{200x}

\label{firstpage}

\maketitle

\begin{abstract}

MERLIN observations of the unusually slow nova V723 Cas are
presented. Nine epochs of 6~cm data between 1996 and 2001 are mapped
showing the initial expansion and brightening of the radio remnant,
the development of structure and the final decline. A radio light
curve is presented and fitted by the standard Hubble flow model for
radio emission from novae in order to determine the values of various
physical parameters for the shell. The model is consistent with the
overall development of the radio emission.  Assuming a distance of
2.39~($\pm$0.38)~kpc and a shell temperature of 17000~K, the model yields values
for expansion velocity of 414~$\pm$~0.1~kms$^{-1}$ and shell mass of
1.13~$\pm$ 0.04~$\times$~10$^{-4}$~M$_{\odot}$. These values are consistent with
those derived from other observations although the ejected masses are
rather higher than theoretical predictions. The structure of the shell
is resolved by MERLIN and shows that the assumption of spherical
symmetry in the standard model is unlikely to be correct.

\end{abstract}


\section{Introduction}

Classical novae are interacting binary systems and the most energetic
type of cataclysmic variable. The central system consists of a white
dwarf primary and a main sequence secondary which fills its Roche
lobe. Hydrogen-rich matter is accreted from the secondary onto the 
white dwarf via an accretion disc. Once the pressure at the base of 
the accreted envelope reaches a critical level, thermonuclear reactions 
begin under degenerate  conditions leading to an explosion
(Starrfield 1989). As a direct result the bolometric luminosity of
the system increases by at least three orders of magnitude over a
timescale of a few days. Observational estimates of the mass of
material ejected are typically $10^{-5}$ -- $10^{-4}$~M$_{\odot}$ with
velocities in the range a few hundred to several
thousand~kms$^{-1}$. Correlations exist between the rate at which the
visual light declines from maximum and the peak absolute magnitude and
the ejection velocity. These are in the sense that the novae which
fade most rapidly (``fast novae'') are intrinsically brightest and
eject material at the highest speeds.

V723 Cas (Nova Cassiopeiae 1995) was discovered at $V=9.2$~mag on 
1995 August 24 by M. Yamamoto (Hirosawa et al 1995) which for the purposes of
this paper will be defined as day zero. The nova reached visual
maximum of $V$=7.1~mag 115 days later on December 17. The light
curve characteristics of V723 Cas display very slow evolution with a
long pre-maximum halt (Hachisu \& Kato 2004) and have led it to be
classified as a very slow nova alongside HR Del and RR Pic (Chochol
\& Pribulla 1997).  Oscillations in the light curve are evident
post-maximum, a feature also seen in the evolution of HR Del. Optical
spectroscopy (Iijima, Rosino \& Della Valle 1998, Munari et al 1996)
prior to maximum light revealed fairly narrow emission components
(full width half maximum FWHM$\sim$ 90~kms$^{-1}$) with P Cygni
absorptions blueshifted $\sim 100$~kms$^{-1}$ relative to the emission
peak. Munari et al (1996) report that post-maximum the emission lines
increase by a factor of several times in intensity and broaden to
FWHM$\sim550$~km~s$^{-1}$ whilst the absorption components remain at a
similar width to pre-maximum. Ground and space-based infrared observations 
(Evans et al 2003) show line profiles with FWHM$\sim 330$~km~s$^{-1}$ and
indicate an ejected mass of $2.6\times 10^{-5}$~M$_\odot$ from the 
Br$\alpha$ line and $4.3\times 10^{-4}$~M$_\odot$ from the free-free 
emission (assuming a distance of 4~kpc).

\begin{table*}
\label{obstable}
\centering
\begin{minipage}{140mm}
\caption{Summary of the MERLIN observations of V723 Cas.}
\begin{tabular}{llllll} 
\hline
Observation &Wavelength &Days since&Fitted Beam          &Fitted Beam  &Flux Density      \\ 
date        &(cm)       &discovery & Size (mas$\times$mas)&PA (degrees) &(mJy)  \\ 
\hline
1996 Dec 13 & 6         & 477      &55.3 $\times$ 44.5   &$-$77.6        &2.3 ($\pm$ 0.5) \\
1997 Jan 25 & 6         & 520      &51.3 $\times$ 44.1   &54.8         &2.4 ($\pm$ 0.4) \\
1998 Mar 04 & 6         & 923      &57.8 $\times$ 44.7   &$-$29.2        &7.0 ($\pm$ 0.2) \\
1998 Apr 06 & 18        & 956      &147 $\times$ 126.3   &$-$29.3        &4.3 ($\pm$ 0.3) \\
1998 Dec 07 & 6         & 1201     &162.5 $\times$ 54    &1.7          &8.2 ($\pm$ 0.5) \\
2000 Feb 27 & 6         & 1648     &124 $\times$ 52.9    &28.9         &7.9 ($\pm$ 0.4) \\
2000 Apr 16 & 6         & 1697     &54.8 $\times$ 35.7   &$-$34.7        &13.5 ($\pm$ 0.9)  \\
2001 Jan 29 & 6         & 1985     &124.5 $\times$ 45.6  &33           &6.3 ($\pm$ 1.1) \\
2001 Jun 03 & 6         & 2110     &139.3 $\times$ 65.3  &15.1         &4.0 ($\pm$ 0.7) \\ 
2001 Oct 26 & 6         & 2255     &46.7 $\times$ 41.3   &$-$89.8       &5.4 ($\pm$ 0.3) \\ 
\hline
\end{tabular} 
\end{minipage} 
\end{table*}

Radio emission in novae typically arises due to free-free
emission from gas at temperatures of approximately 10$^4$~K (Seaquist
1989). Non-thermal components have been detected in a few novae, the
most notable example being GK Per which exhibits a non-thermal ridge
coincident with the south-west portion of the optical shell (Bode, O'Brien 
\& Simpson 2004). This is generally attributed to the collision between the
nova ejecta and the previously shed common-envelope of a pre-nova phase, most
likely following a `born-again' AGB star phase (Dougherty et al 1996). Higher temperature
emission detected in some nova outbursts (e.g. QU Vul, Taylor et
al. 1987; V1974 Cyg, Pavelin et al 1993) suggests that, at least for some novae, 
there is shocked material within the ejected shell.

This paper presents results obtained from ten MERLIN observations of
V723 Cas, nine at a wavelength of 6~cm and one at 18~cm. The 6~cm
epochs produce maps and the shell structure is discussed. Radio flux
measurements lead to a determination of the radio light curve to which
standard models are fitted in order to determine values for the physical
parameters of the nova outburst.

\section{Observations}

V723 Cas was observed using MERLIN between
1996 December 13 and 2001 October 26 resulting in ten epochs of data. 
A summary of the observations can be found in Table \ref{obstable}. The fitted beam is a 2D Gaussian fit to the point spread function for each observation.

The observing wavelength was 6~cm (4994~MHz) for all epochs except
6 April 1998 when it was 18~cm (1658~MHz). The nova was observed in
phase-referencing mode, switching to phase-reference source 0102+511
(J2000) at regular intervals.  The point source OQ208 was used as an
amplitude calibrator and was flux calibrated against 3C286 using the
Baars scale (Baars et al 1977). The data were edited and flagged using the
MERLIN package {\sc dplot} and the flux scale for each epoch was
calibrated using {\sc dproc}. Further calibration, mapping and
analysis was performed using {\sc aips}. Each epoch was mapped to a
$512 \times 512$ pixel image. The phase-calibrator was also mapped
using three passes of self-calibration. Due to the relatively weak
emission from V723 Cas the phase-solutions from the phase calibrator
were used to correct the data as self calibration was not
possible. The data were also reweighted to allow for differences in
the sensitivities of the telescopes in the MERLIN array. 

\section{Results}

\subsection{The 6~cm radio maps}

The nine radio maps at 6~cm are presented in Figure 1.  Natural
weighting of the data was used during mapping to maximise the signal
to noise ratio and the {\sc clean}ing process was terminated when the
total {\sc clean}ed flux reached a maxmimum value. A 50~mas circular 
restoring beam was used to create each map.

\begin{figure*}
\begin{center}
\setlength{\unitlength}{1cm}
\begin{picture}(10,17)
\put(-4,0){\includegraphics{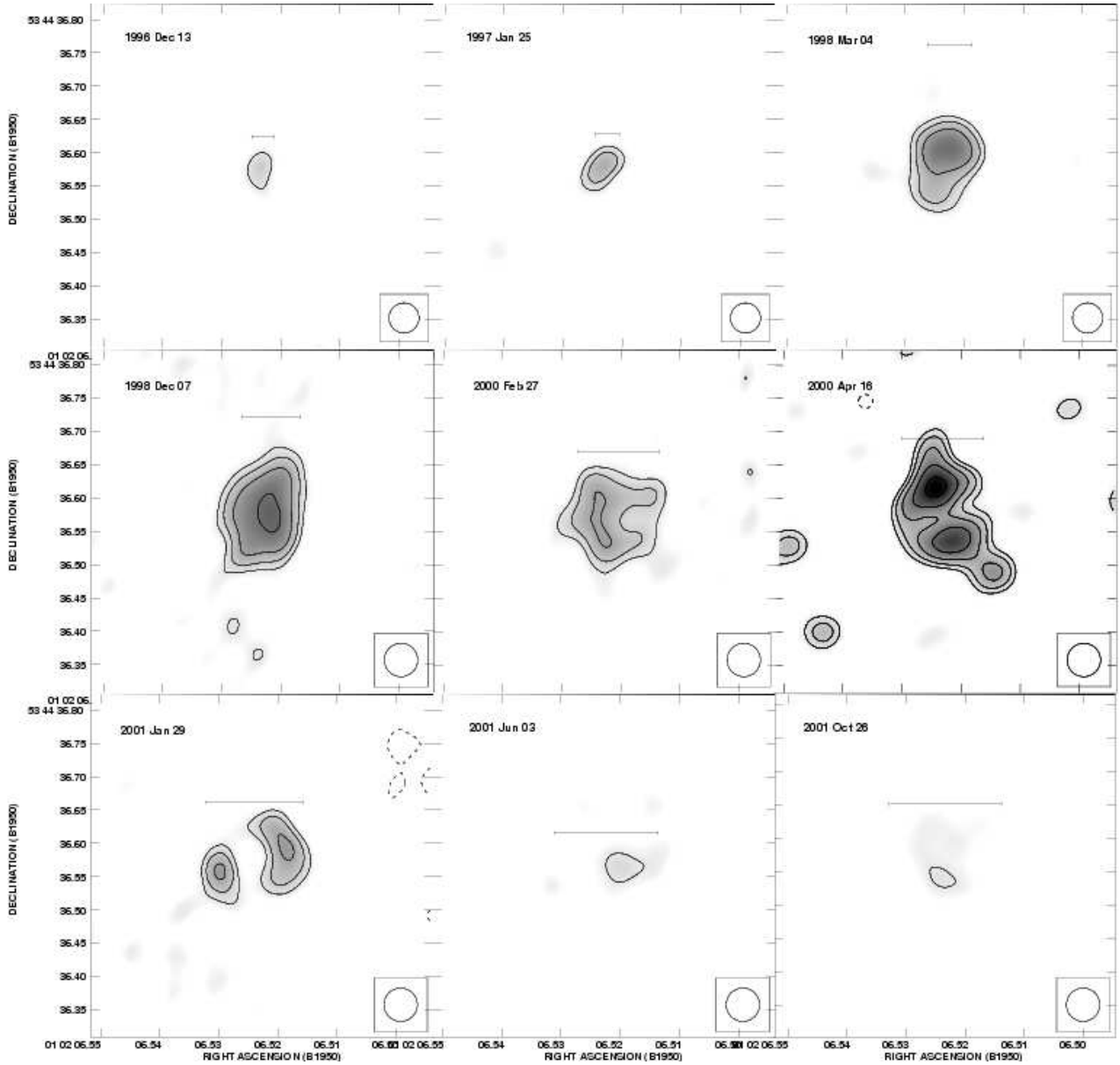}}
\end{picture}
\caption{6~cm MERLIN maps of V723 Cas. The contour levels are
(-3, 3, 3$\sqrt{2}$, 6, 6$\sqrt{2}$, 12, 12$\sqrt{2}$...) $\times$ 0.15~mJy. The greyscale runs
from 3~mJy to 20~mJy. The horizontal bars
show the expected diameter of the shell assuming a particular 
distance and expansion velocity as discussed in the text.}
\end{center}
\label{radiomaps}
\end{figure*}

The December 1996 and January 1997 maps show that the source is
unresolved by MERLIN at this early stage. As the shell expands over
the next few epochs it appears to become extended along a north-south
direction.  Over a year later in February 2000 this north-south axis
is still apparent although it now gives the appearance of being one
side of a partial shell. The April 2000 data are rather noisy (see
section \ref{fluxdensities}) and only the two brightest peaks aligned
with the north-south ridge are believed to be real emission from V723 Cas.  
The next image, taken 9 months
later in January 2001, suggests that the north-south components have
been replaced by ones aligned east-west. This structure then fades
over the remaining two epochs.

\subsection{Radio flux densities}
\label{fluxdensities}

Two methods were used for estimating the total flux density at each
epoch.  Firstly, the total flux density in the region of the source in
the final {\sc clean}ed map was measured using {\sc imstat} and the
uncertainty was determined from the rms variation in an off-source
region of the map. Secondly, the area was re-imaged using {\sc
uvtaper} to weight down the longer baselines. Tapering values of
500~k$\lambda$, 1000~k$\lambda$ and 2000~k$\lambda$ were applied to the
data and the data were also imaged with no tapering.  This is
particularly useful at epochs where the source is resolved since in
these circumstances an interferometer cannot fully sample all the
scale-sizes within the source.  A Gaussian was then fitted to the
source using {\sc imfit} and the total flux density in this Gaussian
and its uncertainty were recorded. The clean depths used for each of
the methods were 1$\sigma$, 2$\sigma$ and 3$\sigma$ and the data were
also imaged by cleaning down to the first negative clean
component. The ten final flux measurements used from the set of 160
produced by these methods were chosen on the basis of the single
imaging method which consistently gave the highest flux density. Using
the two techniques allows an appreciation of the systematic
uncertainties in the fluxes and leads to the estimates presented in
Table \ref{obstable}. Figure 2 shows the resulting
radio light curve for V723 Cas.  The April 2000 observation appears to
have unusually high flux density. Examination of the flux density of
the phase calibrator reveals that it is also suspiciously high and
that the data has poor phase calibration as a result. It is highly
unlikely that the flux density value for this epoch is trustworthy.

\begin{figure*}

\centering
\setlength{\unitlength}{1cm}
\begin{picture}(10,10)
\put(-1,0){\includegraphics{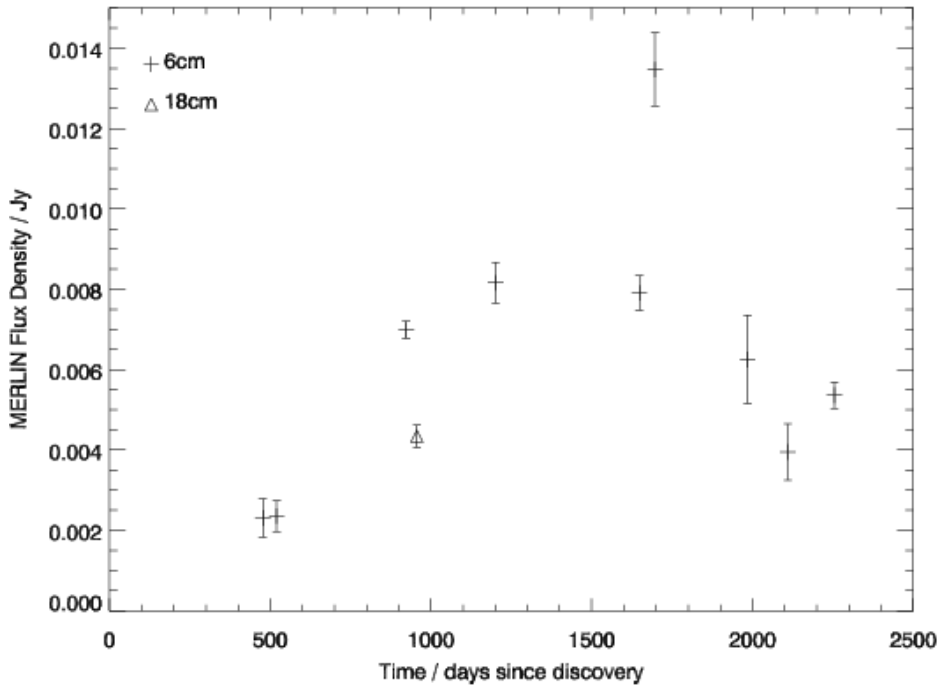}}
\end{picture}
\caption{Radio light curve for V723 Cas. Nine 6~cm epochs and one 18~cm epoch 
are presented. Note the discussion in the text about the anomalously high flux
density at day 1697.}
\label{lightcurve}
\end{figure*}

\section{Interpretation}

\subsection{The radio light curve}

The radio light curve is consistent with the classic 3-phase evolution
for nova light curves (Hjellming 1990). The expanding shell is assumed
to emit via thermal bremsstrahlung. In the earliest stages, Phase I,
the shell is optically thick and the observed flux increases as the
shell expands, the photosphere being coincident with the physical
shell boundary.  The flux is therefore expected to be proportional to
$\nu^{2}t^{2}$, where $\nu$ is the frequency of the radio observation
and $t$ is the time since outburst.  During Phase II the effective
photosphere lags behind the physical shell boundary as the shell
becomes optically thin.  Eventually the shell is entirely optically
thin and, if it is isothermal, the flux declines as
$\nu^{-0.1}t^{-3}$. Hjellming suggests that the flux in Phase
II varies approximately as $\nu^{0.6}t^{-4/3}$.

It is possible to make a simple check on the spectral index of the
emission using the flux densities from the contiguous observations on
1998 March 4 at 6~cm and on 1998 April 6 at 18~cm. This leads to a
spectral index of 0.54 which implies that the turnover into optically
thin emission has begun by this stage. However, by using the 6~cm flux density from March 4 rather than April 6 the spectral index is underestimated. This is consistent with the simple model proposed by Hjellming which predicts a spectral index between 0.6 and 2 for this phase of the evolution.

\subsection{Hubble flow model for radio emission from novae}

By fitting a model for the emission to the light curve of Figure 2 it is possible to estimate the values of a number of
physical parameters such as the mass of the shell and its ejection
velocity.  The standard model employed here is known as the Hubble
flow model (Seaquist and Palimaka 1977; Hjellming et al. 1979). It
assumes an isothermal spherically-symmetric shell resulting from an
instantaneous mass ejection with a linear velocity gradient. Hence as
the shell expands it grows thicker with internal and external
boundaries determined by the assumed range of ejection velocities.

In each case the flux density expected at a given epoch and frequency
is calculated by integrating the radiative transfer equation through
the shell assuming a temperature and distance.  The predicted values
are then compared with the 6~cm observations (excluding the dubious
2000 April observation) and a chi-squared minimisation is used to
estimate the best-fitting values of the parameters.

We estimate the temperature of the radio-emitting shell for use in the
model-fitting by converting the peak flux density in the
optically-thick 1996 December map to an equivalent brightness
temperature of 17000~K.  The distance is taken to be 2.39~($\pm$0.38)~kpc (Chochol
\& Pribulla 1997).

\begin{table}
\begin{center}
\caption{Parameters for the Hubble flow model assuming a shell temperature of 17000~K 
and a distance of 2.39~kpc.}
\begin{tabular}{ll} \hline
Model parameter & Best-fitting value \\ \hline Ejected mass $M_e$ &
1.13 $\pm$ 0.04 $\times$10$^{-4}$~M$_{\odot}$ \\ Outer shell velocity
$v_{1}$ & 414.4 $\pm$ 0.1 km$\,$s$^{-1}$ \\ Inner shell velocity $v_{2}$
& 102 $\pm$ 4 km$\,$s$^{-1}$ \\ \hline
\end{tabular}
\end{center}
\end{table}

Best-fitting values of the model parameters are listed in Table 2. The
reduced chi-squared value is 1.6 indicating a satisfactory fit to the
data.  We discounted the possibility that the fitting routine has
become trapped in a local minimum of the $\chi^2$ hyperspace by 
gridding the $\chi^2$ values at high resolution. 
The uncertainties on the best-fit values for each parameter quoted in Table 2 are
determined from marginalised likelihood distributions of $\chi^{2}$
derived from a gridded region of parameter space around each best-fit
position.  For the Hubble flow model we can deduce from this method
that the ejected mass, the shell velocity and the ratio of inner to
outer velocity are all well constrained and that the likelihood peaks
at the best fit value.

\begin{figure}
\centering
\setlength{\unitlength}{1cm}
\begin{picture}(8,6)
\put(-0.5,-0.5){\includegraphics{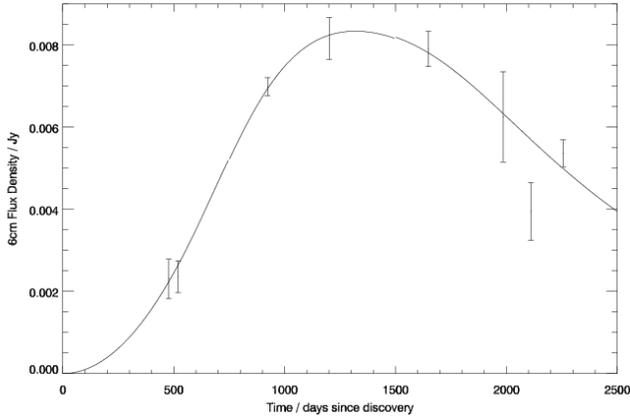}}
\end{picture}
\caption{Best fitting model light curve for the Hubble flow model
overlayed on the flux density measurements from the 6~cm MERLIN observations.}
\label{hubcurve}
\end{figure}

\begin{figure}
\centering
\setlength{\unitlength}{1cm}
\begin{picture}(8,5)
\put(0,-0.3){\includegraphics{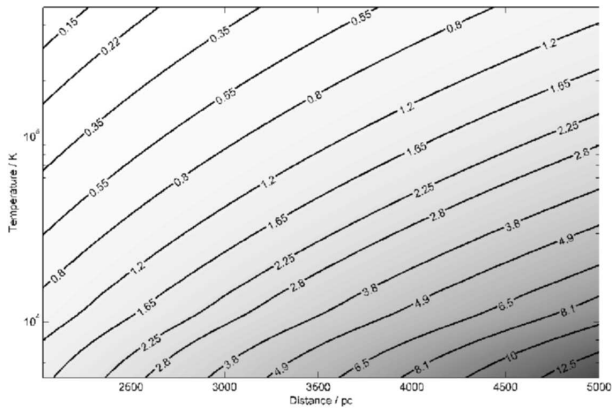}}
\end{picture}
\caption{Best fitting mass contours for a range of temperatures and distances. 
The best fitting mass at any point in the plane is the contour value $\times$~10$^{-4}$~M$_{\odot}$}
\label{masssurf}
\end{figure}

\begin{figure}
\centering
\setlength{\unitlength}{1cm}
\begin{picture}(8,5)
\put(0,-0.3){\includegraphics{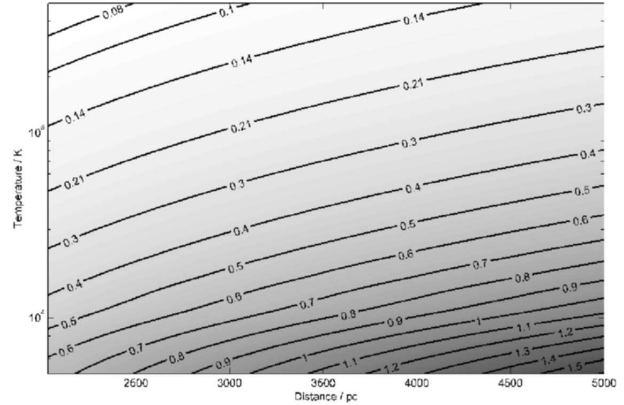}}
\end{picture}
\caption{Best fitting velocity contours for a range of temperatures and distances. 
The best fitting velocity is the contour value $\times$~1000~kms$^{-1}$.}
\label{velsurf}
\end{figure}

Figure \ref{hubcurve} shows the best fitting Hubble flow model light
curve. The best fit for these models depends on both the distance to the nova
and the temperature. 
Figures \ref{masssurf} and \ref{velsurf} show how
the values of mass and velocity vary for different pairs of
temperature and distance. The ratio of inner to outer velocity is fixed
at 0.248 for all these model runs. These plots are the result of 10000
model runs resulting from 100 values each for temperature and distance. 

The estimate of ejected mass derived from the radio lightcurve
modelling of around $1.1\times 10^{-4}$~M$_{\odot}$ is about 25\% of that
obtained from the free-free infrared continuum observations of Evans
et al (2003). However they assume that the distance is 4~kpc. At the
same temperature (17000~K), Figure~4 shows the ejected
mass estimate from the radio emission rises to around $4\times
10^{-4}$~M$_{\odot}$. However, for this distance, assumption of a higher
temperature (Evans et al use a value of $3.2\times 10^{5}$~K) would
reduce the mass estimate again to around $0.8\times
10^{-4}$~M$_{\odot}$. As Evans et al (2003) point out, the
estimates derived observationally are generally about an order of
magnitude greater than those resulting from simulations of the
thermonuclear runaway. 

The higher temperature cited above is thought
to arise from `coronal' gas around the nova. It is unlikely that this
component of the ejecta makes anything other than a negligible
contribution to the radio flux of the nova. This claim is based on the
fact that RS CVn
stars have radio fluxes comparable to that of a nova shell arising
from coronal gas at temperatures in excess of 10$^{8}$~K (Hjellming \&
Gibson, 1980; Kuijpers \&  van der Hulst, 1985). It is
therefore unlikely that coronal gas at 3.2$\times 10^{5}$~K would
significantly affect the results presented in this paper, although inclusion of
this emission is an aspect which would improve the accuracy of future models.

Despite the reasonable fit to the radio light curve the Hubble Flow
model cannot be an entirely accurate representation of the truth. The
smooth mass distribution which the model is based on is misleading as
images of nova shells, including the radio maps presented in this
paper, show that the ejecta are far from uniform. In reality the
material is highly clumped with polar and equatorial features of
enhanced density (e.g. Harman \& O'Brien 2003).
 
A simple statistical treatment for the clumping of the ejecta
has been applied to investigate the effects that a non-uniform
distribution of material may have on the best fitting parameters of
the shell model.  A similar method has been used to investigate
mass-loss from OB stars (Abbott, Bieging \& Churchwell 1981) and from
Wolf-Rayet stars (Nugis, Crowther \& Willis 1998). A correction factor
is applied to the optical depth equations when solving the radiative
transfer through the shell. The correction factor is governed by two
parameters; $x$ is the ratio of the lower density background to the
high density clumps, and $f$ is the filling factor, the fractional
volume occupied by the high density clumps, or alternatively the
fractional length along any line of sight which passes through the
high density material.  In this case, for $x$~$\geq$~0.3 the clumping
changes the parameters by no more than a few percent, however for
strong clumping ($x$~$\sim$~0.01, $f$~$\sim$~0.01) the best fitting
mass can drop to approximately one third of the smooth shell case. The
velocity of the outer shell boundary may increase by $\sim$~7\% for
strongly clumped shells although generally the effect on this
parameter is very small, probably due to it being well constrained
during the optically thick stage. The inner shell velocity increases
by up to a factor of 3 as $f$ and $x$ decrease, suggesting that
increasing the density condensations in this model results in a
thinner shell. A more thorough investigation into clumping in nova
shells has been made by Heywood (2004).

\subsection{Structure of the remnant}

\begin{figure}
\centering
\setlength{\unitlength}{1cm}
\begin{picture}(8,7)
\put(-1.5,8){\includegraphics{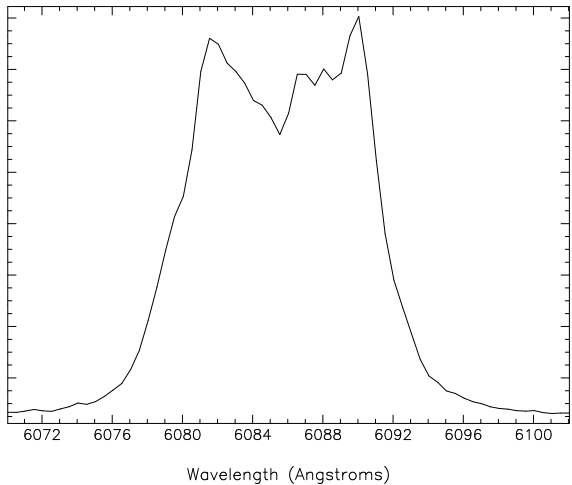}}
\end{picture}
\caption{[Fe{\sc VII}] line for V723 Cas from O'Brien et al (in prep). The flux scale is relative and the data were taken on 1999 February 4 with the Intermediate Dispersion Spectrometer on the Isaac Newton Telescope.}
\label{lineprofile}
\end{figure}

The apparent angular size of the expanding remnant can be estimated
from the expansion velocity and distance.  Expansion velocities are
measured spectroscopically from the [Fe{\sc vii}] (6087~\AA) line
profile presented in Figure~\ref{lineprofile} (from O'Brien et al, in preparation). 
This forbidden line is chosen because
it is optically thin and has a well-defined double-peaked
structure. The expansion velocity can be measured from either peak to
peak separation, full width at half maximum (FWHM) or the full width at zero
intensity (FWZI).  In this case these are 210~kms$^{-1}$, 290~kms$^{-1}$ and  
660~kms$^{-1}$ respectively. In the analysis below 
the peak to peak value is used as it probably represents the 
velocity of the majority of the ejected mass.
Acceleration is assumed to be negligible for this calculation. 

The radio lightcurve modelling resulted in velocities ranging from 
102 to 414~kms$^{-1}$ (inner and outer boundaries of the Hubble flow model) 
which is generally consistent with the velocities 
derived spectroscopically. For example, the early spectroscopy (Munari 
et al 1996, Iijima et al 1998) showed P Cygni absorptions at velocities 
about 100 km/s blueward of the emission peak, consistent with the lower 
velocities derived from the radio models. 
The expansion velocity derived from the average FWHM of the infrared 
emission lines (Evans et al 2003) of 332$\pm$17~kms$^{-1}$ is 
slightly larger than the value derived above of 290~kms$^{-1}$.

The assumption that all material is ejected instantaneously with a
linear velocity gradient is another feature of the Hubble-flow model
which may be too simplistic. Spectroscopic analysis of nova shells
indicates the presence of many velocity components. In some cases it
appears that a relatively slow, high-mass component is embedded in a
fast, tenuous wind. In this case, spectra of V723 Cas from 1996 August show a
component at very high velocity of $\sim$1500~kms$^{-1}$ in the P
Cygni profiles of both infrared and optical lines (Evans et al 2003,
O'Brien et al, in preparation).  It should also be noted that the
spherically symmetric wind-like Hubble-flow model cannot be entirely
correct as it produces flat-topped line profiles in the optically-thin
stage rather than the double-peaked profile seen in
Figure~6.  Future versions of these models will
attempt to take these features into account.

Using a distance of 2.39~($\pm$0.38)~kpc and an expansion time measured
from the date of discovery (as the exact date of the outburst is
unknown) the horizontal bars shown in Figure 1 represent the predicted
angular size of the remnant at each epoch.
Care must be taken when comparing the spectroscopically calculated
angular sizes with the extent of the contours on the radio images. In
later epochs the outer regions of the shell have faded below
detectable levels so comparing the size of the outer contours with the
bar is futile. This comparison is valid in the optically thick phase
where the predicted angular sizes are generally in good agreement
with the radio contours.

It would be interesting if the radio contours could be used to determine the distance to the nova from expansion parallax. This would be difficult since the radio contours do not accurately represent the size of the object beyond the optically thick regime. A detailed model to fit the radio contours would have to be implemented in order to estimate the distance. This will be addressed in a paper to follow.

An interesting effect is seen between the 2000 April and 2001 January
observations. The north-south structure appears to be replaced by an
east-west alignment of the brightest peaks. A possible explanation for
the switching of the dominant emission axis involved optical depth
effects caused by the ejecta being non-uniform.
Initially the ejecta are optically thick so in the early radio
observations only the photosphere is visible. The brightness of the
source is dependent only upon its temperature and angular size. As the
remnant expands the density decreases and the radio photosphere begins
to lag behind the physical shell boundary. This is the point where
structure in the shell may become evident in the radio
observations. However, the higher density regions scattered throughout
the shell would remain optically thick for longer than the low density
regions and would appear as bright peaks in the radio maps. As the
expansion continues it is possible that the bright peaks that appear
in one observation may become optically thin and be replaced by high
density regions at different positions in the shell giving the
impression that the axis of the dominant emission regions has
rotated. Nova shells are known to exhibit coherent regions of increased density such as polar caps and tropical rings (e.g. FH Ser -- Gill \& O'Brien, 2000) and these features would be ideally placed for the apparent north-south / east-west switching observed between epochs.

This apparent rotation is an effect that has been seen in other interferometric
observations of classical novae e.g. V1974 Cyg (Eyres et al 1996) and V705
Cas (Eyres et al 2000). The 6~cm maps for V1974 Cyg show north--south
structure developing for the first five epochs only to be replaced by
east--west features in the final epoch. The corresponding 18~cm maps
show the same effect but occuring at a later time. Since the emission
at 18~cm will become optically thin at a later time than the 6~cm
emission this may support the hypothesis that an optical depth effect
is responsible.

However interpretation of radio interferometry maps is complicated
since emission on certain scales can be `resolved out' and the
incomplete sampling of the $uv$ plane means that the beam shape
significantly influences the image. Structures on the scales seen in
these maps are particularly prone to these sort of instrumental
effects.  Further progress in their interpretation will only be gained
by simulated interferometric images of model shells which has been carried out and will be
presented in a paper to follow.

\section{Conclusion}

Radio emission from V723 Cas has been detected at both 6~cm and
18~cm 477 days after discovery of the nova and has been monitored for 
a further 1778 days. The radio maps show non-uniform structure with a N-S ridge feature
that develops over time and splits into two peaks which then switch to
E-W alignment in the next map. The radio light curve is consistent
with the evolution of other novae of faster speed classes.

A simple Hubble-flow model is applied to the radio flux measurements
yielding values for expansion velocity (414.4$\pm$0.1)~kms$^{-1}$ and
shell mass (1.13$\pm$0.04)~$\times 10^{-4}$~M$_{\odot}$ for a distance of 
2.39~($\pm$0.38)~kpc and temperature 17000~K. Although
the value for ejected mass is consistent with previous estimates for classical novae
there is still significant discrepancy between the mass predicted
from the theory of the outburst (e.g. Starrfield, Truran \& Sparks 2000) 
and the mass derived
from modelling of observations such as those presented in this
paper. Expansion velocities are similar to those derived from spectroscopic 
observations although the latter show that certain assumptions of the standard 
Hubble-flow model for radio emission are naive.

Asphericity of the shell and clumping of the ejected material are
evident in many novae and future models should take these factors into
account. The radio maps show how the morphology of the shell evolves and
further understanding of this will only be gained when simulated model
images are convolved with the $uv$ distribution of the array and true
synthetic images are produced. This will also provide insight into how
nova shells will be seen by the next generation of interferometry
networks such as $e$-MERLIN.

\section*{Acknowledgements}

MERLIN is operated as a UK National Facility by the University of Manchester, 
Jodrell Bank Observatory, on behalf of the Particle Physics and Astronomy 
Research Council (PPARC). IH and MFB are grateful to PPARC for provision of a 
postgraduate studentship and Senior Fellowship respectively. IH also wishes to thank 
Peter Blake of the Department of Computer Science at the University of Manchester for his
data handling experience.

\bsp

\label{lastpage}

\end{document}